\documentclass[twocolumn,showpacs,preprintnumbers,amsmath,amssymb]{revtex4} 

\usepackage{graphicx}
\usepackage{dcolumn}
\usepackage{bm}
\usepackage{appendix}

\newcommand\lnn{\mbox{ln}}

\begin{document}

\preprint{APS/123-QED}

\title{E=mc$^2$ and the negative pressure of Dark Energy.}

\author{B. Castaing}
 \email{bernard.castaing627@laposte.net}
 
\affiliation{Laboratoire des \'ecoulements g\'eophysiques et industriels\\
Domaine Universitaire, CS 40700, 38058 Grenoble Cedex 9, France}

\date{\today}

\begin{abstract}
According to recent observations, the Dark Energy would represent 70\% of the content of our Universe. The most popular way to account for this Dark Energy make use of the Cosmological Constant introduced by Einstein. However, some uncertainty remain about the theoretical value this constant should have. Alternate approaches have to introduce a fluid whose pressure would be negative. In this paper, we present unusual particles which could form a gas with positive thermodynamic pressure, while presenting all the gravitational characteristics of Dark Energy.  
\end{abstract}



\maketitle

\section{Introduction. }
\label{int}

According to the present knowledge \cite{Palanque}, ordinary mass, mainly made of baryons in stars or interstellar gases, account only for 4\% of the total content of our Universe. Another 23\% is made of yet unobserved particles, but behaves classically in the sense that its total mass is conserved. This part is called the Dark Matter because it presents no interaction with photons or ordinary particles, and is observable only through its gravitational effects.

The remaining 73\% is called Dark Energy. Its existence is required to explain a change of rate in the expansion of the Universe, which occurred approximately 6Gyrs in the past. Before this date, the expansion was decelerating, now it is accelerating. This implies that the total corresponding mass increases when the Universe expands. The most popular way to account for this Dark Energy is to introduce a Cosmological Constant in the Einstein's equations of General Relativity. However, many questions remain about the pertinence of this constant, and the value it should have \cite{cosmos}.

In this context, where many efforts have been made and brillant ideas have been raised, the goal of this paper is modest. It aims at drawing the attention on an alternate possibility for the Dark Energy, which should be taken seriously. It concerns the dispersion relation of vacuum excitations ({\it i.e.} particles) in the neighborhood of the Planck wave length. The unusual particles proposed are first presented. Then the properties of a gas of such particles is examined, and their incidence on the Universe behavior.

Due to the apparent conflict between the ideas presented here and common beliefs, we prefer to proceed by steps. In section \ref{part}, we present the simplest model for these unusual particles, and calculate their contribution to the inertial mass density. In section \ref{rho}, we discuss the interpretation of this mass density, and show that it must be considered also as a gravitational mass. In section \ref{evolution}, we determine the relation of this mass density with the volume of the Universe, and show that it implies that the total corresponding mass increases as this volume expands. It yield us to a new interpretation of the apparent negative pressure of Dark Energy. However, the way this mass increases when the volume expands do not exactly correspond to the observations concerning the Dark Energy. Put in other words, the parameter $w$ of the adiabatic equation of state \cite{Carroll} would be $w=-2$, while recent observations \cite{Planck}  imply that it should be close to $w=-1$. In section \ref{w1}, we show how a slight modification of the model allows to take it into account. We group a few remarks in section \ref{rem}, including a crude model suggesting that the Dark Matter could as well be explained in this frame. We conclude in section \ref{conc}.

\section{Unusual particles. }
\label{part}

To the author knowledge, all particles previously proposed to account for the missing mass in the Universe have a dispersion relation whose minimum is for a wave vector $\vec{k}=0$ (or momentum  $\hbar\vec{k}=0$). The proposition here is that a particle exists (which we could call a darkon) for which this minimum is at a non zero wave vector $k_p$. The corresponding dispersion relation would thus locally be:
\begin{equation}
\hbar\omega(k)=\Delta+\frac{\hbar^2(k-k_p)^2}{2m},
\label{disp}
\end{equation}
where $\Delta=mc^2$, and $c$ is the light velocity. In Appendix \ref{lag}, we propose a Lagrangian which can yield to such a dispersion relation.

Note that this is plausible. A condensed matter example of similar (quasi-)particles is given by rotons in superfluid $^4$He \cite{Maris}, which can act as a model for the present proposition. In the case of  $^4$He, the existence of such a minimum in the dispersion relation of excitations is related to the tendency of helium to crystallize, {\it i.e.} to the spontaneous appearance of a mass density wave. Helium thus presents quite a softness for the corresponding wave length. In the case of interest here, let us note that some theories \cite{Bertone} postulate that supplementary dimensions are wrapped as tubes. This tendency could also exist for our physical dimensions, inducing a softness of vacuum for wave length of order the size of these tubes, thus presumably the Planck length:
\begin{equation}
k_p \simeq \sqrt{c^3/\hbar G} \simeq 10^{35}m^{-1},
\end{equation}
where $G$ is the gravitation constant.

We have some arguments for assuming the following interesting properties for these particles.
\begin{itemize}
\item i) They have vanishing interactions with ordinary particles. This should be the case, as they appear as very small objects compared to the de Broglie wave length of these ordinary particles.
\item ii) They remain in a thermal equilibrium distribution during the expansion of the Universe, either because of strong interactions between them, or because the expansion is a reversible process for a gas of such particles.
\item iii) As soon as the temperature of this gas is small compared to the gap $\Delta$, their number is conserved. Indeed, they cannot decay as ordinary particles while conserving both energy and momentum. Even the process where two darkons would give a bunch of ordinary particles has a very reduced possible phase space, as the two darkons must have almost opposite momenta.
\end{itemize}

 Property iii), which means that the total number of these particles remains constant, is very important in the following. A detailed examination of property ii) would need highly speculative hypotheses on the interaction between darkons, which are out of the scope of this paper. At this stage, we prefer to refer to the behavior of rotons in the model system of superfluid $^4$He. In this system, below a temperature of 1K, a decoupling is predicted and observed between phonons and rotons. On a reduced temperature range, a sound  is observed in this roton gas, whose velocity shows that the number of rotons is conserved \cite{Khalatnikov,Maris,Castaing}.

Under the above hypotheses,  the population of the mode of wave vector $\vec{k}$ at equilibrium, $n_{\vec{k}}$, must maximise the entropy, while conserving the number of particles, the total energy and the total momentum. When varying the population $n_{\vec{k}}$ by $\delta n_{\vec{k}}$, the condition of constant number writes:
\begin{equation}
\sum_{\vec{k}}\delta n_{\vec{k}}=0.
\label{num}
\end{equation}

As for the conditions of constant energy and momentum, they respectively write:
\begin{equation}
\sum_{\vec{k}}\hbar\omega(k)\delta n_{\vec{k}}=0,
\label{eng}
\end{equation}
and
\begin{equation}
\sum_{\vec{k}}\hbar\vec{k}\delta n_{\vec{k}}=0.
\label{mom}
\end{equation} 

Under the Maxwellian approximation, the condition of maximum entropy writes:
\begin{equation}
\sum_{\vec{k}}\lnn (n_{\vec{k}}) \delta n_{\vec{k}}=0.
\label{ent}
\end{equation}

The method of Lagrange multipliers then yield to:
\begin{equation}
\lnn (n_{\vec{k}})=\alpha-\beta\hbar\omega(k)+\vec{\gamma}\hbar\vec{k},
\label{max1}
\end{equation}
where $\alpha$, $\beta=1/k_BT$, and the three components of $\vec{\gamma}$ are constants, and $T$ is the temperature. The population $n_{\vec{k}}$ thus can be written:
\begin{equation}
n_{\vec{k}}=\exp(-(\hbar\omega(k)-\hbar\vec{k}\vec{v}-\mu)/k_BT),
\label{max}
\end{equation}
where $\vec{v}$ is the drift velocity of the gas formed by these particles,  $\mu$ their chemical potential , and $T$ the temperature. The value of $\mu$ is given by the condition that the total number of darkons remains constant.

The Maxwellian approximation, which avoids to decide between bosonic or fermionic character for these new particles, is valid as long as $k_BT<<\Delta-\mu$. This approximation will be justified in section \ref{rem}. Let us simply remark that it would need a huge particle density to be wrong (a particle each Planck length).

The particle density $n$ writes:
\begin{eqnarray}
n&=&\int \frac{d^3\vec{k}}{8\pi^3}n_{\vec{k}} \nonumber \\
&=&\frac{k_p^2}{2\pi^2}\frac{\sqrt{2\pi mk_BT}}{\hbar}\exp(-\frac{\Delta-\mu}{k_BT}).
\end{eqnarray}

The momentum density of the gas is the sum of the contributions of each mode $\vec{k}$. Up to first order in $v$, it gives, per unit volume \cite{nm}:
\begin{equation}
\rho\vec{v}=\int \frac{d^3\vec{k}}{8\pi^3} \hbar\vec{k} \frac{\hbar\vec{k}\vec{v}}{k_BT}\exp(-(\hbar\omega(k)-\mu)/k_BT).
\label{ro1}
\end{equation}

The coefficient $\rho$ plays the role of an inertial mass density. As $k_BT<\Delta=mc^2<\hbar k_pc$, most of the populated modes are close to $k_p$. Then:
\begin{equation}
\rho=n\frac{\hbar^2k_p^2}{3k_BT}.
\label{ro}
\end{equation}

Note that, within this approximation, $\rho$ does not depend on $v$, and the absolute referential, implicitly assumed in equation \ref{max}, does not matter.

\section{Interpretation of this specific mass.}
\label{rho}

The interpretation of this specific mass $\rho$ is crucial. Certainly, it is not an ordinary mass, as can be seen from the dependency with temperature. However, it is really an inertial mass, with a corresponding kinetic energy  $\rho v^2/2$. According to the Einstein equivalence  between inertial and gravitational mass, we also must consider it as a gravitational mass, contributing to gravity. Let us show it by the following argument.

Consider a volume element $\delta V$ of the darkons gas, later designed as 1, in the gravitational field of a mass $M$, later designed as 2. As the considered velocities are small, the Newtonian approximation is valid. If gravity can be resumed to a geometrical effect, the trajectory of 1 will be the same as for any object of the same initial position and velocity (falling bodies law). Its acceleration will thus be:
\begin{equation}
\frac{d\vec{v}}{dt}=\frac{GM\vec{R}}{R^3},
\end{equation}
where $\vec{R}$ is the vector from 1 to 2. The momentum rate of 1 being, to first order in $v$:
\begin{equation}
\rho\delta V\frac{d\vec{v}}{dt},
\end{equation}
the force that 2 exerts on 1 is:
\begin{equation}
\vec{F}_{2/1}=\rho\delta V\frac{d\vec{v}}{dt}=\frac{GM\rho\delta V\vec{R}}{R^3}.
\end{equation}

The action and reaction law, {\it i.e.} the conservation of momentum, implies that 1 exerts on 2 the force:
\begin{equation}
\vec{F}_{1/2}=-\frac{GM\rho\delta V\vec{R}}{R^3}.
\end{equation}

It means that the darkons gas creates a gravity field whose potential is:
\begin{equation}
\phi(\vec{r})=\int\frac{G\rho(\vec{r'})}{|\vec{r}-\vec{r'} |}d^3\vec{r'}.
\end{equation}

Always in the Newtonian approximation, the contribution of darkons to the diagonal temporal component of the curvature tensor is \cite{Landau}:
\begin{equation}
\mathcal{R}_0^0=\frac{1}{c^2}\Delta\phi=\frac{4\pi G}{c^2}\rho.
\label{R00}
\end{equation}

According to the Einstein equation \cite{Landau}:
\begin{equation}
\mathcal{R}_j^i=\frac{8\pi G}{c^4}(\mathcal{T}_j^i-\frac{1}{2}\mathcal{T}\delta_j^i),
\end{equation}
where $\mathcal{T}_j^i$ is the energy momentum tensor, we should have:
\begin{equation}
\mathcal{R}_0^0=\frac{4\pi G}{c^4}\varepsilon',
\end{equation}
$\varepsilon'$ being the energy density of the gas. But $\varepsilon'$ is of order $n\Delta$. Equation \ref{R00} shows that we must replace $\varepsilon'$ by $\varepsilon=\rho c^2$.

This is not to say that $\varepsilon$ is the real energy density of the gas. We rather think that the discrepancy between $\varepsilon'$ and $\varepsilon$ reflects neglected terms in the Einstein equation. Indeed, these equations can be inferred from the sole hypothesis that the action only contains the components of the metric tensor and their first derivatives \cite{Landau}. This is clearly a large wave length approximation, not valid at scales comparable to the Planck one. A sufficiently important population of darkons, with their extremely short wave length, could make higher derivative terms become significant. 

We show in Appendix \ref{field} in a different context, that of superfluids, how a similar hypothesis limiting the order of derivatives of the density present in the action, yield to non dispersive phonons as the only possible excitation. On the opposite, introducing derivatives of the density in the action, we can obtain a dispersion relation similar to the roton-darkon one.

\section{Evolution of $\rho$ with the volume.}
\label{evolution}

The above argument shows that the usual Einstein equations can be used if, in the energy-momentum tensor, the energy density corresponding to the darkons gas is replaced by $\varepsilon_d=\rho c^2$, which we can call the density of "inertial energy", much larger than its thermodynamic energy density, of order $n\Delta$.

As can be seen from the equation \ref{ro} the density $\rho$ depends on the temperature. We have thus to determine how the temperature behaves at constant number of particles (property iii)) when the gas expands with the Universe.

The gas pressure is given by the perfect gas law:
\begin{equation}
P=nk_BT.
\end{equation}

As the gas evolution is presumably at constant entropy $S$, the expression of this entropy will be useful. One can obtain it from the Gibbs grand potentiel $\Omega=-PV$:
\begin{equation}
S=-\frac{\partial\Omega}{\partial T})_{\mu,V}.
\end{equation}

The main contribution comes from the exponential term in $n$:
\begin{equation}
S \simeq nVk_B\frac{\Delta-\mu}{k_BT}=nVk_B(\ln(n/\sqrt{T})+cst).
\label{ent}
\end{equation}

At constant number $N=nV$, the adiabatic equation thus writes $V^2T=cst$, or $PV^3=cst$, or $T^3/P^2=cst$. It can also be written $\rho/V=cst$, or $\rho^2T=cst$, or $\rho^3P=cst$. 

Considering the "inertial energy":
\begin{equation}
\mathcal{I}=\rho c^2V,
\end{equation}
we can define an "inertial pressure":
\begin{equation}
\varpi=-\left(\frac{\partial\mathcal{I}}{\partial V}\right)_S=-2\rho c^2.
\end{equation}

 This is consistent with the often made statement that the Dark Energy has "a strong negative pressure". Moreover, the corresponding $w$ parameter of the equation of state would be $w=-2$ \cite{Carroll}. Note however that $\mathcal{I}$ is not the thermodynamic energy and $\varpi$ is not the thermodynamic pressure. The real thermodynamic pressure is $P$, which is positive, and respects all the stability criterions of usual gases.
 
 Nevertheless, $w=-2$  is excluded by recent observations, particularly the results of the Planck mission \cite{Planck}. It  would mean that the equivalent mass density increases with the expansion of the Universe, while it is observed to remain constant, which corresponds to $w=-1$. In the next section \ref{w1}, we show how the model can be modified to result in a gas with $w=-1$.

\vspace*{5mm}

\section{A gas with $w=-1$.}
\label{w1}

Indeed, it is possible to modify the previous model in such a way that $w=-1$. For this purpose, we have to consider the following dispersion relation instead of equation \ref{disp}:
\begin{equation}
\hbar\omega(k)=\Delta+\hbar c|k-k_p|.
\label{disp2}
\end{equation}

Again, a Lagrangian as proposed in Appendix \ref{lag} can yield to such a dispersion relation. It must be confessed that, here, we cannot refer to another experimental example of such a dispersion relation. A Dirac like dispersion generally corresponds to a zero gap. However, if we accept such a hypothesis, most of what was derived concerning the previous case is also valid now, with a few exceptions. The pressure $P$ is always given by the perfect gas law $P=nk_BT$, but $n$ is now given by:
\begin{equation}
n=\frac{k_p^2}{\pi^2}\frac{k_BT}{\hbar c}\exp(-\frac{\Delta-\mu}{k_BT}).
\end{equation}

The inertial mass density is, here also, related to $n$ by:
\begin{equation}
\rho=n\frac{\hbar^2k_p^2}{3k_BT},
\label{ro2}
\end{equation}
again not related to the thermodynamic energy density. The entropy now writes:
\begin{equation}
S \simeq nVk_B\frac{\Delta-\mu}{k_BT}=nVk_B(\ln(n/T)+cst).
\label{ent2}
\end{equation}

At constant number $N=nV$, the adiabatic equation thus writes $VT=cst$, or $PV^2=cst$, or $T^2/P=cst$. It can also be written very simply $\rho=cst$, as $n/T$ is $\rho$, within a constant. 

Considering again the "inertial energy", which plays the role of energy in the Einstein gravitation equations:
\begin{equation}
\mathcal{I}=\rho c^2V,
\end{equation}
and the "inertial pressure":
\begin{equation}
\varpi=-\left(\frac{\partial\mathcal{I}}{\partial V}\right)_S=-\rho c^2,
\end{equation}
which means that $w=-1$. This is sufficient to ensure that the acceleration of the expansion of the Universe occurs at the right time, as it is observed. We explicit the reasoning in Appendix \ref{exp}.

Even if fully developing the question of the dynamics of such a gas is not in the scope of this paper, note that this dynamics should be surprising. For instance, there cannot be a gravitational instability in such a gas, whatever its size: under compression, the gravitational mass decreases so rapidly that the gravitational energy decreases, despite the reduced radius. Also, the pressure increases more rapidly than for ordinary three dimensional gases ($\gamma$, the ratio of constant pressure to constant volume specific heats, is 2). Inertial compressions could occur, due to non uniformity of the velocity $\vec{v}$, at the points where div$\vec{v}<0$, but the decrease of the total mass close to these points, and the strong increase of pressure, together with the opposite effects where div$\vec{v}>0$, constitute an efficient restoring force. Obviously, detailed simulations are necessary to confirm, but all these mechanisms tend to make the density of the darkons gas uniform. Again, this is an expected property of the Dark Energy.

\vspace*{5mm}

\section{Final remarks.}
\label{rem}

In this section, we shall first justify the Maxwellian approximation, giving an estimation for the ratio $(\Delta-\mu)/k_BT$. Then, we shall give arguments which suggest that the present model could account not only of the Dark Energy, but also of the Dark Matter.

In the second model ($w=-1$), the particle density writes:
\begin{equation}
n=\frac{k_p^2}{\pi^2}\frac{k_BT}{\hbar c}\exp(-\frac{\Delta-\mu}{k_BT}).
\end{equation}

Assimilating the corresponding mass density to that of the Dark Energy, we have:
\begin{equation}
\rho=n\frac{\hbar^2k_p^2}{3k_BT}\simeq 0.7\,10^{-26}kgm^{-3}.
\end{equation}

The two above equations yield to:
\begin{equation}
\exp(-\frac{\Delta-\mu}{k_BT})=3\rho\frac{\hbar c\pi^2}{\hbar^2k_p^4}\simeq 10^{-122},
\end{equation}
or:
\begin{equation}
\frac{\Delta-\mu}{k_BT} \simeq 280,
\end{equation}
which largely justifies the Maxwellian approximation.

As for the Dark Matter, let us remark that the gap $\Delta$ could be smaller around structures, as is the case around defects in superfluids. Then, as soon as the temperature is smaller than the variation of the gap, darkons will be trapped, and will accumulate around these structures. Once trapped, the temperature of these particles will not change, neither their effective mass. These structures with their trapped particles can thus model the Dark Matter.

The number $N$ of darkons which are not yet trapped should follow a dynamics like:
 \begin{equation}
 \frac{1}{N}\frac{dN}{dt}=-n_S\sigma_Sc,
 \label{dyN}
 \end{equation}
where $n_S$ is the density of structures, and $\sigma_S$ their cross section for darkons. Assuming that the structures remain identical, and are simply diluted by the expansion, their density at any time verifies:
\begin{equation}
n_S=n_{pS}\frac{V_p}{V},
\label{ns}
 \end{equation}
 where $n_{pS}$ is the present density of structures, and $V_p$ the present volume of the Universe. To make this model more concrete, we thus need to have the time evolution of $V$, and to identify these hypothetical structures with observable objects.
 
 For the time evolution of $V$, we shall make the crude model that it follows from a constant mass in an almost flat, isotropic Universe. Then, using, for instance, equation \ref{hcst} with $B=0$:
 \begin{equation}
  \frac{1}{V}\frac{dV}{dt}=\frac{K}{\sqrt{V}} \mbox{\hspace{5mm} and \hspace{5mm}} V \propto t^2,
  \label{vt}
 \end{equation}
 where the origin of time is taken when $V=0$. Using equations \ref{dyN}, \ref{ns}, and \ref{vt}, we have:
 \begin{equation}
  \frac{1}{N}\frac{dN}{dt}=-n_{pS}\sigma_Sc\frac{t^2_p}{t^2} \mbox{\hspace{1mm} and \hspace{1mm}} \ln (\frac{N_i}{N_p}) \simeq n_{pS}\sigma_Sc\frac{t^2_p}{t_i},
 \end{equation}
 where $t_p$ is the present time, $t_i$ the time where the trapping begins, $N_i$ the total number of darkons, $N_p$ the present number of untrapped ones.
 
 As we shall see, today, most of darkons should be trapped in the structures, and contribute to the Dark Matter. Then:
  \begin{equation}
  \frac{N_i}{N_p}=\frac{\rho_{DM}T_i}{\rho_{DE}T_p}=\frac{\rho_{DM}V_p}{\rho_{DE}V_i}=\frac{\rho_{DM}t_p^2}{\rho_{DE}t_i^2},
   \end{equation}
   where $\rho_{DM}$ (resp. $\rho_{DE}$) is the present mass density of Dark Matter (resp. Dark Energy), $T_p$, the present temperature of untrapped darkons, $T_i$ their temperature when the trapping begins, which is also the temperature of the trapped ones. We have thus the following equation for $t_i$:
 \begin{equation}
 \frac{t_i}{t_p} \ln (\frac{\rho_{DM}t_p^2}{\rho_{DE}t_i^2})=n_{pS}\sigma_Sct_p.
 \label{ti}
 \end{equation}
  
  To go further, we shall assimilate the positions of the structures to the positions of the galaxies, considering, as is often made \cite{Combes}, that they acted as seeds for the galaxies ({\it via} the mass of the trapped darkons). To estimate $n_{pS}$ and $\sigma_S$, we shall use the known numbers for our own galaxy, the Milky Way. It should contain $3 \, 10^{11}$ stars, for a mass of approximately $M_{MW} \simeq 3 \, 10^{41}$kg, and a radius of $R_{MW} \simeq 10^{21}$m. The present average mass density corresponding to luminous objects (like stars) should be $\rho_{\ell} \simeq 4 \, 10^{-29}$kgm$^{-3}$ \cite{Palanque}. We thus estimate $n_{pS}$ as:
 \begin{equation}
 n_{pS}=\frac{\rho_{\ell}}{M_{MW}} \simeq 10^{-70}\mbox{m}^{-3}.
 \end{equation}
 
 As for $\sigma_S$, it is probably a fraction of $R_{MW}^2$:
 \begin{equation}
 \sigma_S = \alpha R_{MW}^2,
 \end{equation}
  with $\alpha<1$. We are conscious that the Milky Way could not be the average galaxy, but smaller galaxies should also have less stars, and the estimated value of $n_{pS}\sigma_S$ could be poorly sensitive to the typical galaxy considered.
  
  All these estimations give the following approximate solution for equation \ref{ti}:
  \begin{equation}
   \frac{t_i}{t_p} \simeq \alpha^{\zeta}\times 10^{-3},
   \end{equation}
   where $\zeta \simeq 7/6$, to take into account the logarithmic correction. In the extreme case where $\alpha \simeq 1$, this gives $t_i \simeq 10$Myrs, {\it i.e.} before the re-ionization of the Universe (500Myrs). It is remarquable that such a crude model give a plausible result.
  
  \section{Conclusion.}
  \label{conc}
  
  To conclude, we want to stress the twofold nature of the ideas discussed above.
  
  On the one hand, assuming the existence of darkons, {\it i.e.} particles whose energy minimum is at a finite momentum, gives an alternate interpretation of the Dark Energy, not referring to a Cosmological Constant, and avoiding the notion of negative pressure. The apparent negative pressure would simply come from a misinterpretation of the gravitational mass of this gas, whose real, thermodynamical pressure is positive. As discussed in the previous section \ref{rem}, the same model can potentially take into account the Dark Matter as well.
 
 Questions remain, however. The interaction between these darkons should be clarified: the justification of our thermodynamic treatment rely mainly to a simple analogy with a model system. Going further would need to study the dynamics of the darkons gas, and to precise some parameters as the gap $\Delta$ and the temperature $T$.

On the other hand, the discussion raises interesting questions about very high wavenumber excitations (Planck scale), namely their possible signature at our scale, and the necessity of additional terms in the Einstein equations they reveal. Indeed, either by the concrete consequences of their existence, or by the virtual implications of it, the very short wavelength excitations could have an important role to play in our vision of the Universe.

\acknowledgments
The author expresses all his gratitude to Professor Fran\c coise Combes for her splendid paper in "Reflets de la Physique" \cite{Combes} which initiated this work. I also thank her and Jean-Loup Puget for numerous interesting remarks and discussions.

\appendix

\section{Expansion evolution.}
\label{exp}

We explicit here the consequences of the existence of darkons, with the dispersion relation \ref{disp2}, on the Universe expansion. A starting point can be the isotropic expansion model of Friedmann-Lema\^itre-Robertson-Walker (FLRW),  as described for instance in the Landau - Lifchitz Theoretical Physics course \cite{Landau}. In this model, the Universe grows as a hypersphere of radius $a$, whose volume is  $2\pi^2a^3$. Its evolution is given by (equation 108,7 \cite{Landau}):
\begin{equation}
\frac{da}{cdt}=\sqrt{\frac{8\pi G}{3c^4}\varepsilon a^2-1}.
\label{evol}
\end{equation}

$\varepsilon$ is the mass density multiplied by $c^2$. This mass density has two components. On the one hand, we have the ordinary mass density, which results from the conservation of the total classical mass. We shall include in it the dark matter, even if the exact nature of the corresponding particles is unknown. It gives a term $A/a^3$ in $\varepsilon$.

On the other hand, there is the density  $\rho$ discussed above (the $w=-1$ case). In an adiabatic expansion, it remains constant, thus a term $B$ in $\varepsilon$. This is similar to what would result with a finite value of the Cosmological Constant, which would also give a constant term.

The equation \ref{evol} then becomes:
\begin{equation}
\frac{da}{cdt}=\sqrt{\frac{8\pi G}{3c^4}(Ba^2+A/a)-1}.
\label{evol3}
\end{equation}

The term:
\begin{equation}
\frac{8\pi G}{3c^4}(Ba^2+A/a),
\end{equation}
has its minimum for $a_m=(A/2B)^{1/3}$.

Considering that today, the density of Dark Energy is approximately 3 times the density of Matter:
\begin{equation}
B=3\frac{A}{a^3} \mbox{\hspace{5mm} thus \hspace{5mm}} a=6^{1/3}a_m.
\end{equation}

The instantaneous Hubble "constant" $H$ is:
\begin{equation}
H=\frac{da}{adt}=\sqrt{\frac{8\pi G}{3c^2}(B+A/a^3)-c^2/a^2}.
\label{hcst}
\end{equation}

In the limit of a flat Universe, which seems to be the case \cite{Planck}, the last term under the square root is negligible. Integrating $1/H$ {\it versus} $\ln a$, we find that the expansion acceleration would have begun $6.6 \times10^9$ years before present, in fair agreement with observations.

\section{A Lagrangian for the darkons.}
\label{lag}

We want here to stress that a Lagrangian can be proposed which yield, in limiting cases, to the dispersion relations \ref{disp} and \ref{disp2}. Consider indeed the following Lagrangian:
\begin{equation}
\mathcal{L}=p_ov-mc^2\sqrt{1-v^2/c^2}-L_o,
\label{la}
 \end{equation}
 where $v$ is the modulus of the velocity $\vec{v}$, and $p_o$ and $L_o$ are constants. The components of the momentum are given by:
\begin{equation}
p_i=\frac{\partial \mathcal{L}}{\partial v_i}= p_o\frac{v_i}{v}+\frac{mv_i}{\sqrt{1-v^2/c^2}}.
\label{mom}
\end{equation}

The Hamiltonian is:
\begin{equation}
\mathcal{H}=p_iv_i-\mathcal{L}=L_o+\frac{mc^2}{\sqrt{1-v^2/c^2}}.
\label{ham1}
\end{equation}

After some algebra, equation \ref{mom} gives the following relation between $p$ and $v$:
\begin{equation}
\frac{m^4v^4}{(1\!-\!v^2/c^2)^2}\!-\!2(p^2\!+\!p_o^2)\frac{m^2v^2}{(1\!-\!v^2/c^2)}\!+\!(p^2\!-\!p_o^2)^2\!=\!0.
\label{rel}
\end{equation}

Let us set:
\begin{equation}
x=\frac{1}{(1-v^2/c^2)} \mbox{\hspace{5mm} thus \hspace{5mm}} v^2=c^2(1-\frac{1}{x}).
\end{equation}

Equation \ref{rel} gives the following second order equation for $x$:
\begin{equation}
x^2\!-\!2x(\frac{p^2\!+\!p_o^2}{m^2c^2}\!+\!1)\!+\!(\frac{(p^2\!-\!p_o^2)^2}{m^4c^4}\!-\!2\frac{p^2\!+\!p_o^2}{m^2c^2}\!+\!1)\!=\!0.
\end{equation}

The only interesting root of this equation is:
\begin{equation}
x_-=\frac{(p-p_o)^2}{m^2c^2}+1.
\end{equation}

The other root would give a high energy branch without interest in our context. Thus the hamiltonian can be written:
\begin{equation}
\mathcal{H}=L_o+mc^2\sqrt{\frac{(p-p_o)^2}{m^2c^2}+1}.
\label{ham2}
\end{equation}
 
 Using the equivalences $p=\hbar k$ and $\mathcal{H}=\hbar \omega$, equation \ref{ham2} reduces to the dispersion relations \ref{disp} and \ref{disp2} in the two following limits:
\begin{itemize}
\item i) When $L_o=0$ and $|p-p_o|<<mc$, we find equation  \ref{disp}. The gap is $\Delta=mc^2$.
\item ii) When $m$ is vanishingly small, and $|p-p_o|>>mc$, we find equation  \ref{disp2}. The gap is $\Delta=L_o+mc^2$.
\end{itemize}

\section{A different, while instructive, context.}
\label{field}

As remarked above,  the Einstein equations can be inferred from the sole hypothesis that the action only contains the components of the metric tensor and their first derivatives \cite{Landau}. Let us examine the consequences of a similar hypothesis in a different context, that of a superfluid, {\it i.e.} a fluid without dissipation.

We shall follow the presentation of Uwaha and Nozi\`eres \cite{Uwaha}. Let us thus consider a fluid whose potential energy only depends on its density: $U(\rho)$, with a minimum for $\rho_o$. Its Lagrangian is the difference between the kinetic energy $\rho v^2/2$, and this potential energy, integrated over the whole space. The action must be minimized, within the constraint that the mass is conserved:
\begin{equation}
 \partial_t\rho+\partial_i(\rho v_i)=0.
\label{masse}
\end{equation}

Following the Lagrange multipliers method, this is equivalent to minimize without restriction the effective action:
\begin{equation}
 \int dt\int d^3r\,\{\rho \frac{v^2}{2}-U(\rho)-\lambda(x,t)[
\partial_t\rho+\partial_i(\rho v_i)]\}.
\end{equation}

Varying the action,  $\rho \rightarrow \rho+\delta\rho$ and $v_i \rightarrow v_i+\delta v_i$ add five terms under the sum:
\begin{equation}
 L_1\delta\rho+L_{2i}\partial_i\delta\rho+L_{3}\partial_t\delta\rho+L_{4i}\delta
v_i+L_5\partial_i\delta v_i,
\end{equation}
with
\begin{eqnarray}
 L_1 & = & \frac{1}{2} v^2-\mu-\lambda\partial_i v_i ,\\
 L_{2i} & = & -\lambda v_i ,\\
 L_3 & = & -\lambda ,\\
 L_{4i} & = & \rho v_i-\lambda\partial_i\rho ,\\
 L_5 & = & -\lambda\rho,
\end{eqnarray}
where $\mu=\partial U/\partial\rho$. $\mu$ is the chemical potential, and $d\mu=dP/\rho$, where $P$ is the pressure.

But, integrating by parts:
\begin{equation}
 \int\!dt\int\!d^3rL_{2i}\partial_i\delta\rho= \int\!dt\int\!d^3r(-\partial_iL_{2i})\delta\rho.
\end{equation}

Transforming the same way the third and fifth terms give two groups of equations. $L_{4i}-\partial_i L_5=0$ yield to $v_i=-\partial_i\lambda$. The velocity is a gradient. The equation of motion is then:
\begin{equation}
 \partial_t v_i +\partial_i(\mu+\frac{1}{2} v^2)=0.
\label{mvt1}
\end{equation}

This is the Euler equation, if we remark that $\partial_i(v^2/2)-v_j\partial_jv_i=v_j(\partial_iv_j-\partial_jv_i)=0$ as the velocity is a gradient. The linearized version of equation \ref{mvt1} can be written:
\begin{equation}
 \partial_t v_i +\frac{d^2U}{d\rho^2}\partial_i\rho=0,
 \label{lmvt1}
\end{equation}
and the conservation of mass, equation \ref{masse}, becomes:
\begin{equation}
 \partial_t\rho+\rho_o\partial_i( v_i)=0.
\label{lmasse}
\end{equation}

The two equations \ref{lmvt1} and \ref{lmasse} yield to:
\begin{equation}
 \partial^2_{t^2} \rho -\rho_o\frac{d^2U}{d\rho^2}\partial_i\partial_i\rho=0,
\end{equation}
which is the equation of propagation of sound of velocity $c_s$:
\begin{equation}
 c_s^2=\rho_o\frac{d^2U}{d\rho^2}.
\end{equation}

The dispersion relation is $\omega^2-c_s^2k^2=0$. The absence of dispersion for the associated phonons is the consequence of the assumed scale invariance. 

Let us now allow the potential energy of the fluid to depend on derivatives of the density, for instance as:
 \begin{equation}
 U(\rho)-\frac{\kappa_1}{2}\partial_i\rho\partial_i\rho +\frac{\kappa_2}{2}\partial^2_{ij}\rho\partial^2_{ij}\rho,
\end{equation}
where $\kappa_1$ and $\kappa_2$ are two positive constants. It means that the fluid is softer for variations of density at a scale of order $\ell=\sqrt{\kappa_2/\kappa_1}$. The velocity is always a gradient: $v_i=-\partial_i\lambda$. The equation of motion becomes:
\begin{equation}
 \partial_t v_i +\partial_i(\mu+\kappa_1\Delta\rho+\kappa_2\Delta\Delta\rho+\frac{1}{2} v^2)=0,
\label{mvt2}
\end{equation}
where $\Delta=\partial_i\partial_i$ is here the Laplacian differential operator. The linearized version of equation \ref{mvt2} can be written:
\begin{equation}
 \partial_t v_i +\frac{d^2U}{d\rho^2}\partial_i\rho+\kappa_1\partial_i\Delta\rho+\kappa_2\partial_i\Delta\Delta\rho=0,
\end{equation}
which, together with the conservation of mass, equation \ref{lmasse}, gives:
\begin{equation}
 \partial^2_{t^2} \rho -\rho_o\frac{d^2U}{d\rho^2}\Delta\rho -\rho_o\kappa_1\Delta\Delta\rho -\rho_o\kappa_2\Delta\Delta\Delta\rho=0.
\end{equation}

The dispersion relation is :
\begin{equation}
\omega^2-k^2(c_s^2 -\rho_o\kappa_1k^2+\rho_o\kappa_2k^4)=0.
\end{equation}

If:
\begin{equation}
3\kappa_2c_s^2<\rho_o\kappa_1^2<4\kappa_2c_s^2,
\end{equation}
this dispersion relation has a minimum, of the roton-darkon type, close to:
\begin{equation}
k_m=\sqrt{\frac{\kappa_1}{2\kappa_2}}.
\end{equation}

In the same way, allowing high order derivatives of the components of the metric tensor in the action could result in (and is necessary for having) a roton-darkon like minimum, at least in the dispersion relation of gravity waves.


\begin{thebibliography}{99}
\bibitem{nm} With the dispersion relation \ref{disp}, and $k_p=0$ (classical particles), equation \ref{ro1} would give $\rho=nm$, as it must be.
\end{thebibliography}
\end{document}